\documentclass[12pt]{article}

\usepackage{amsmath}
\usepackage{graphicx}
\usepackage{subfig}
\usepackage{amssymb}

\begin{document}

\begin{titlepage}

\title{Simulating neurobiological localization of acoustic signals based on temporal and volumetric differentiations}
\author{Nikesh S. Dattani}
\date{October 27, 2008}

\maketitle

\begin{center}
Department of Systems Design Engineering, University of Waterloo \\ Waterloo, Ontario~
 N2L 3G1, Canada
\end{center}

\begin{abstract}
The localization of sound sources by the human brain is computationally simulated from a neurobiological perspective.  The simulation includes the neural representation of temporal differences in acoustic signals between the ipsilateral and contralateral ears for constant sound intensities (angular localization), and of volumetric differences in acoustic signals for constant azimuthal angles (radial localization).  The transmission of the original acoustic signal from the environment, through each significant stage of intermediate neurons, to the primary auditory cortex, is also simulated.  The errors that human brains make in attempting to localize sounds in evolutionarily uncommon environments (such as when one ear is in water and one ear is in air) are then mathematically predicted.  A basic overview of the physiology behind sound localization in the brain is also provided. 
\end{abstract}

\end{titlepage}

\tableofcontents

\newpage

\section{Introduction}

\qquad The capacity of vertebrates to localize sound sources has been preserved over numerous generations of evolution due to the significant benefits it provides regarding the detection of predators.  The brain's ability to detect microsecond-long temporal differences in signals, and to use just this miniscule article of information for calculating the locations of a sound source to a precision of a few degrees, is undoubtedly one of the most remarkable facilities it possesses.

  Learning how this fascinating mechanism operates from a neurobiological point of view has a number of promising applications to biomedical engineering, such as designing aids for patients with damaged cells in the nuclei of the superior olivary complex.  
  
  Additionally, gaining knowledge regarding the brain's mechanism for performing this task may provide insights into improving current synthetic sound localizers, which may someday lead to more practical and clever robots than those that we have designed thus far.  Likewise, methods in acoustical engineering may draw upon such knowledge of our brain to allow technologies such as surround sound speaker systems to portray more realistic manifestations of sounds.

	Numerical experiments on a computer may also provide us with information that may be too expensive or time-consuming to extract out of physical experiments with live mammals.	For example, \textit{Homo sapiens} especially, are much better at locating their distance away from a sound source (\textit{distance localization}), and their angle with respect to the sound source in a 2-dimensional plane at ground level (\textit{azimuthal localization}), than at estimating the elevations of such sources (\textit{vertical localization)}.  By constructing mathematical models of the brain's function that are able to reproduce available experimental results regarding distance localization and azimuthal localization, we may extrapolate to observe how our model performs at the task of vertical localization.  We can then perform numerical experiments to better understand what may improve our vertical localization abilities, with the hope to one day be able to construct devices that allow pilots and air traffic controllers to better detect altitudes of sounds sources.  
	
\section{2-Dimensional Representation: Equidistant-Isovolumentric Signals}

	\qquad I start by modeling the brain's mechanism of representing temporal differences between the left ear and the right ear's detection of an acoustic signal.  I will use the term ``detector'' to denote the human subject that is detecting these acoustic signals.  Here, the signals are kept at a fixed amplitude (they are isovolumetric), and they remain equidistant from the detector's ears, while the sound source's angle with respect to the azimuth\footnote{a straight line passing perpendicularly through the detector's coronal plane} is varied from 0 to 2$\pi$ radians. See Fig. 1 below.\footnote{All angles reported herein will be measured clockwise from the azimuth and will be reported in radians.}
	
	\begin{figure}[h!]
	\centering		
	\includegraphics[scale=0.5]{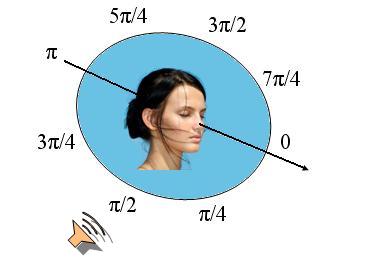}
	\caption{A detector placed in the center of a 2-dimensional plane.  The angle between the speaker displayed and the azimuth (0 radians) measures $\frac{\pi}{2}$ radians.}
	\end{figure}

	The term ``2-Dimensional Representation'' used in the title dictates that the representation discussed in this section is of the position of a sound source in the 2-D plane intersecting the detector's ears.\footnote{It should be noted that since there is no radial variance, the representation itself will be manifested in the form of a 1-D scalar (the angle between the sound source and the azimuth), but since these angles contain information about both the $x-$ and $y-coordinates$ of the source in a 2-D plane, the representation will be called ``2-Dimensional.''  This distinction was made to avoid confusion with the following section.}

	This model uses the Neural Engineering Framework (NEF) developed by Chris Eliasmith and Charles H. Anderson\cite{EliasmithBook}. It begins with a survey of what is known about the system (System Description). 

\subsection{System Description}

\subsubsection{Connectivity and Functional Functional Description}

\qquad Sounds that reach the ear are transduced by hair cells, which create action potentials in neurons of the anteroventral cochlear nucleus.  These action potentials are conveyed, via fibers of cranial nerve VIII, to neurons of each medial superior olive\footnote{There are two medial superior olives: one for the left ear and one for the right.}, which is where processing of temporal differentiations in acoustic signals begins \cite{2000Kandel}.  See Fig. 2 below.   

	\begin{figure}[h!]
	\centering		
	\includegraphics[width = \textwidth]{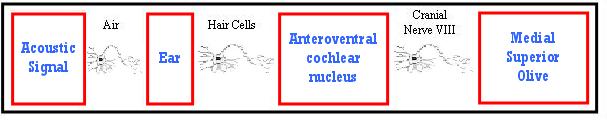}
	\caption{A display of the transmission of acoustic information from the environment to the location where processing of temporal differentiations occurs.}
	\end{figure}

The neurons of the medial superior olive do not reach the threshold required for excitement unless they are stimulated by \textit{both} (left and right) of the incoming neurons from the respective anteroventral cochlear nuclei simultaneously.  But the lengths of the axons of the contralateral\footnote{When referring to a specific side of the body (whether left or right), the \textit{contralateral} side is the opposite side (whether left or right),  while the \textit{ipsilateral} side is the original side in question.} cochlear nucleus vary in length so that \textit{only one} particular axon sends the action potential to its respective receiver in the medial superior olive in time for its stimulus to coincide with that of the ipsilateral axon.  Since the contralateral ear receives the acoustic signal \textit{later} than the ipsilateral ear, and since this delay depends on the source's angle to the azimuth, different neurons of the medial superior olive will be activated for each possible location of the source\footnote{As far as this discussion is concerned, the phrase \textit{each possible location} only refers to locations in one half of the plane, since for example, the time delay associated with a source being 15 degrees from the azimuth, will be the exact same as the time delay associated with the source being 165 degrees from the azimuth.  The brain actually has other ways of detecting whether a signal is coming from the anterior or posterior, but this discussion is omitted for now.}.  For example, in Fig. 3, if there was a big time delay between the left and right ears receiving the signal (i.e. the source's angle to the azimuth was close to 90 degrees), neuron 8 or 9 would probably be activated, since the long time delay due to the temporal differentiation would have to be negated by a shorter time delay from the axon (i.e. a shorter axon length).  

	\begin{figure}[h!]
	\centering		
	\includegraphics[height = 0.65 \textheight , width = 0.5 \textwidth] {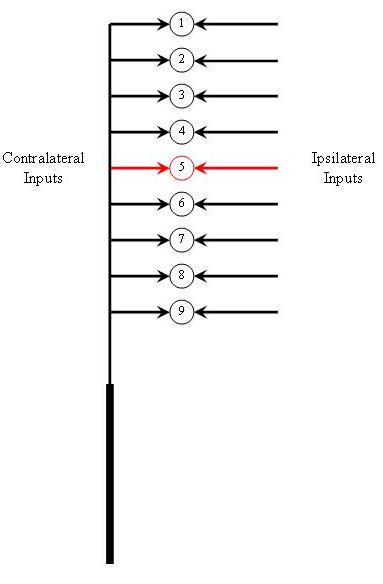}
	\caption{A display of the transmission of acoustic information from the environment to the location where processing of temporal differentiations occurs.  Although only one box for the medial superior olive is shown, signals are actually sent to both the left and right medial superior olives from each ear.  The chronological direction of flow in this chart is denoted by the neurons between each box (from dendrites to axons in all cases).}
	\end{figure}

In Fig. 3, neuron 5 is denoted to be activated, due to the fact that the stimulus probably came from either 0 degrees or 180 degrees from the azimuth.  Of course, in the actual brain there are many more neurons than depicted in the diagram (each representing a particular possible location of the sound source).  

\subsubsection{Tuning Curves}

\qquad In accordance with the above description, each neuron of the medial superior olive is most sensitive to a particular range of possible time differences.  Consequently, each neuron will have two specific ``preferred locations'' for the sound source (remembering that there are two possible locations for the sound source in Fig. 1 that will result in each particular length of time delay - one in the half-plane containing the direction of the azimuth, and one in the half-plane containing the negative of that direction).

The firing rate of a particular neuron will be highest when the sound source is positioned in one of these two ``preferred locations'' and would \textit{ideally} be zero when the sound source is in any other location.  However, it is indeed possible that the firing rate will be non-zero for certain sound locations near enough to the preferred location.  Referring to the example in Fig. 3 for instance, the signal from the contralateral side can reach the medial superior olive soon enough for neuron 6 to fire, or late enough for neuron 4 to fire.

Consequently, the tuning curves have been modeled as narrow Lorentzian curves with peak firing rates chosen between 114/sec and 268/sec\footnote{This is because neurons on the medial superior olive have been recorded with firing rates in this range.\cite{1969Goldberg}}, and reflected to be symmetric about $\frac{\pi}{2}$ radians.  Fig. 4 shows the tuning curve for one of these neurons randomly selected from an ensemble of 4000 neurons - it's maximal firing rate is approximately 170 spikes/second and it's preferred directions are approximately 3.28 radians and 6.14 radians from the azimuth.  Notice how the preferred directions are symmetric about $\frac{\pi}{2}$ radians, and the abscissa of the plot is labeled with $\frac{\pi}{2}$ radians in the centre.

	\begin{figure}[h!]
	\centering		
	\includegraphics[height = 0.4 \textheight , width = 0.95 \textwidth]{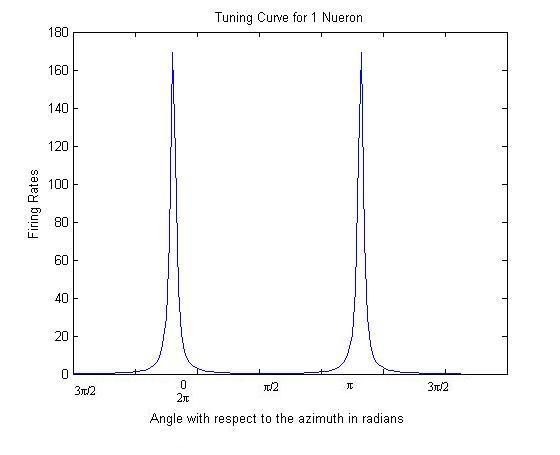}
	\caption{The tuning curve for one neuron chosen out of an ensemble of 4000, showing it's maximal firing rate and it's preferred directions.}
	\end{figure}
  
Fig. 5 displays the tuning curves of 50 neurons chosen out of the ensemble of 4000, and Fig. 6 displays the entire ensemble of neurons.  The preferred locations are randomly chosen between 0 and $2\pi$ radians, with (HWHM)s\footnote{Half-widths at half-heights} randomly chosen between 1 and 2.5 degrees\footnote{This is based on the range of precision with which the neurons of the medial superior olive tend to estimate sound locations based on coincidence detection.  See \cite{2000Kandel} for a reference.} and with amplitudes randomly chosen between 114 and 268.

\begin{figure}[h!]
	\centering		
	\includegraphics[height = 0.41\textheight , width = \textwidth]{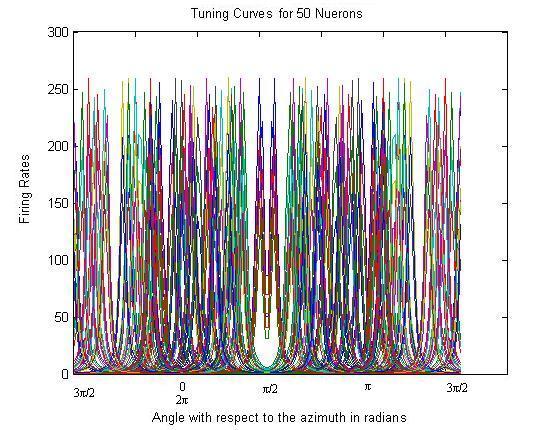}
	\caption{The tuning curves for 50 neurons chosen out of an ensemble of 4000.}
\end{figure}

\begin{figure}[h!]
	\centering		
	\includegraphics[height = 0.41\textheight , width = \textwidth]{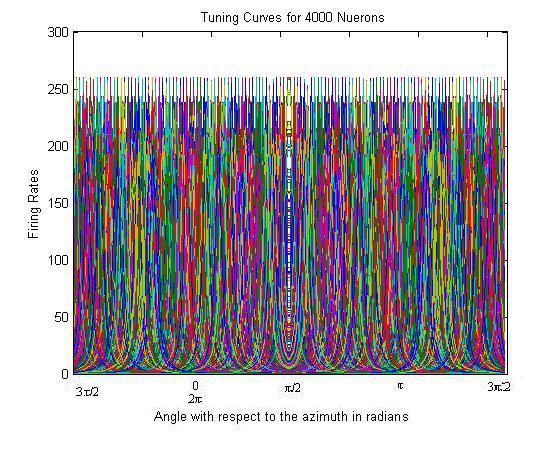}
	\caption{The tuning curves for all 4000 neurons in the ensemble.}
\end{figure}  
 
In Fig. 6 we can see that with 4000 neurons, almost every possible sound source location has a corresponding neuron with maximal firing rate at that location.  Of course, it would be impossible to have a separate neuron for each of the infinite possible sound source locations between 0 and 2$\pi$ radians - the small separation between the peaks of the curves in Fig. 6 leads to the fact that the human brain can only locate sound sources to within a few degrees of precision\cite{2000Kandel}.

\subsubsection{Variable Specifications and Mathematical Description of the System Funciton}

\qquad For this simulation, the variable $x$ will denote the angle that the sound source makes with the azimuth.  The encoding of this variable by each individual neuron $(i)$ will be expressed by the Lorentzian neural response functions (as depicted in Figs. 4-6):

\begin{equation}
a_i(x) = \frac{1}{\pi}\frac{\frac{1}{2}\Gamma}{(x-x_0)^2 + (\frac{1}{2}\Gamma)^2} ~~~~~~~\textrm{for}~~ \frac{\pi}{2} \le x \le \frac{3\pi}{2} ~~~~
\end{equation}

\begin{equation}
a_i(x) = \textrm{fliplr(}\frac{1}{\pi}\frac{\frac{1}{2}\Gamma}{(x-x_0)^2 + (\frac{1}{2}\Gamma)^2} \textrm{) ~~ for} ~~ \frac{3\pi}{2} \le x \le 2\pi ~~ \textrm{and} ~~ 0 \le x \le \frac{\pi}{2} ~~~~
\end{equation}

\vspace{3mm}
Here, $\frac{2}{\pi\Gamma}$ represents the peaks of the Lorentzians and $x_0$ represents the preferred direction for a particular neuron.  The neurons whose preferred sound source locations are to the \textit{right} of the human detector when he or she is facing forward (i.e. between 0 and $\frac{3\pi}{2}$ radians) are labeled herein as `on' neurons whereas those most sensitive to sound source locations on the \textit{left} are labeled as `off' neurons.

\subsection{Design Specification}

\subsubsection{Range}

\qquad The range for the variable $x$ described above is 0 to 2$\pi$ radians.  These values are \textbf{\textit{not}} normalized to range from 0 to 1.  Rather, they are kept as angles measured clockwise from the azimuth in radians.  The purpose of choosing this convention will become clear in subsequent sections of this paper.

\subsubsection{Signal-to-Noise Ratios (SNR)}

\qquad Since neurons carry about 3 bits of information per spike\cite{EliasmithBook}.  As mentioned in Ref. \cite{EliasmithBook}, this means that the coding allows approximately 10\% error, and thus, we can assume that the neurons encode information wtih a signal-to-noise ratio (SNR) of approximately 10:1.  A 10\% error corresponds to a variance of $\sigma^2 = 0.1$ if the distribution of signals is even.  Accordingly, I will assume that the noise is independent and Gaussian distributed with a mean of zero and a variance of 0.1 for each neuron in the model.\footnote{More information regarding the transmission of signals by individual neurons is presented in Ref. \cite{1997Rieke}.}  

\subsubsection{Precision}

\qquad For the precision of the neural representations, it is difficult to predict how precise this model will be before it is implemented, and therefore, I will consider determining the precision of the representation as one of the \textit{purposes} of this model.  In the implementation section, the relationship between the number of neurons and the precision of the representation will be explored. 

\subsection{Implementation}

\subsubsection{Decoding Rules}

\qquad The decoders used in this section are ``optimal linear decoders.''  The decoding can be summarized by the following formula:

\begin{equation}
\hat{x} = \displaystyle\sum_{i=1}^N (a_i(x) + \eta _i)\phi _i
\end{equation}

Here, $\hat{x}$ is the estimation of the signal $x$, $N$ is the number of neurons, $i$ is an index indicating the description of a particular neuron, $a_i$ is the firing rate of neuron $i$, $\eta_i$ is the noise associated with neuron $i$, and $\phi_i$ are the optimal linear decoders given below.

\begin{equation}
\boldmath{\phi} = \boldmath{\Gamma}^{-1} \boldmath{\Upsilon}
\end{equation}

where, each element of $\boldmath{\Gamma}$ is given by:

\begin{equation}
\Gamma _{ij} = \int_0^{2\pi} a_i(x) a_j(x) dx + \sigma^2 \delta_{ij}
\end{equation}

and, each element of $\boldmath{\Upsilon}$ is given by:

\begin{equation}
\Upsilon _j = \int_0^{2\pi} a_j(x) x dx.
\end{equation}

A detailed derivation of the above expressions is demonstrated in Ref. \cite{EliasmithBook}.  

\subsubsection{Results}

\qquad The first implementation of this model was in decoding the signal $x$ using 4000 neurons.  Fig. 7 shows the decoded estimate $\hat{x}$ and the original signal $x$ over the range of $x$.  The root mean square deviation was 0.011.  The difference between the decoded estimate and the original signal is plotted over the range of $x$ in Fig. 8 (notice the change in the scale of the ordinate axis). 

	\begin{figure}[h!]
	\centering		
	\includegraphics[width = \textwidth , height = 0.4 \textheight]{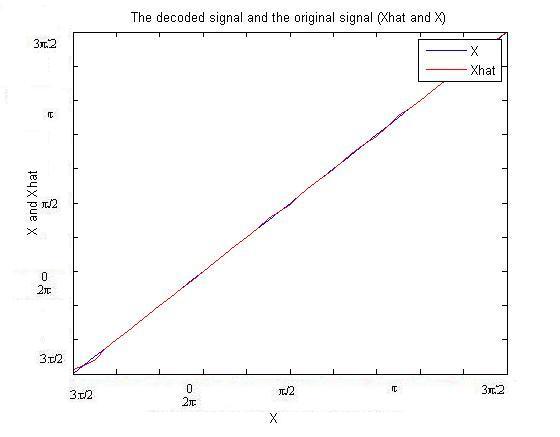}
	\caption{Decoded estimate $\hat{x}$ and original signal $x$ over the range of $x$.}
	\end{figure}

	\begin{figure}[h!]
	\centering		
	\includegraphics[width = \textwidth, height = 0.4 \textheight]{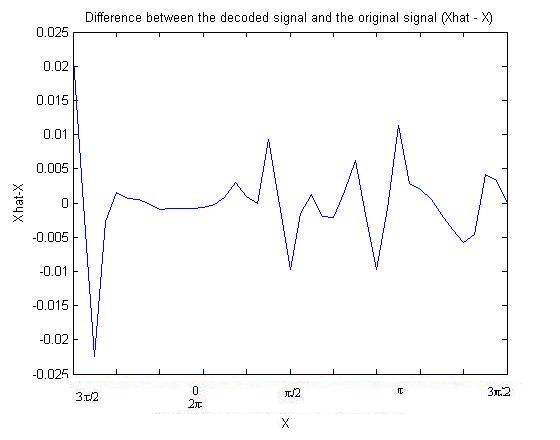}
	\caption{Differences between the decoded estimate and the original signal ($\hat{x} - x$) over the range of $x$.}
	\end{figure}
  
\subsubsection{Numerical Experiments}

\qquad The next item for which this model was implemented was numerical experimentation.  The first of these numerical tests was implemented with the objective to determine the relationship between the number of neurons and the error due to noise and static distortion.  The relationship for the error due to noise should decrease as $\frac{1}{N}$ (where $N$ is the number of neurons) according to previous results \cite{1992Snippe}\cite{1988Paradiso}\cite{EliasmithBook}.  Fig. 9 verifies this expected result, and Fig. 10 displays the error due to static distortion against the number of neurons.

	\begin{figure}[h!]
	\centering		
	\includegraphics[width = \textwidth , height = 0.4 \textheight]{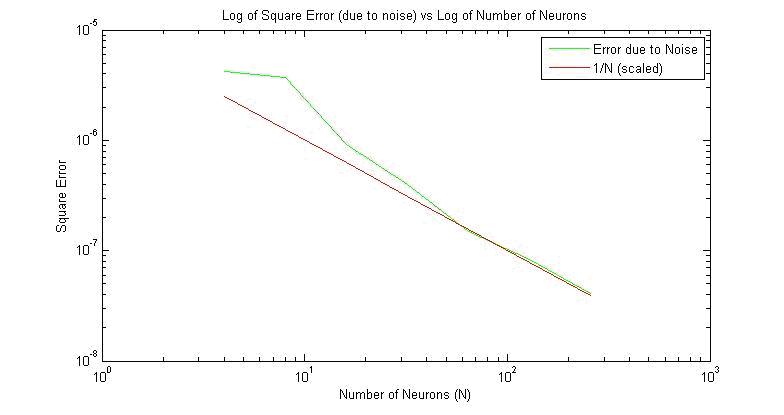}
	\caption{Error due to noise with increasing number of neurons.}
	\end{figure}

	\begin{figure}[h!]
	\centering		
	\includegraphics[width = \textwidth, height = 0.4 \textheight]{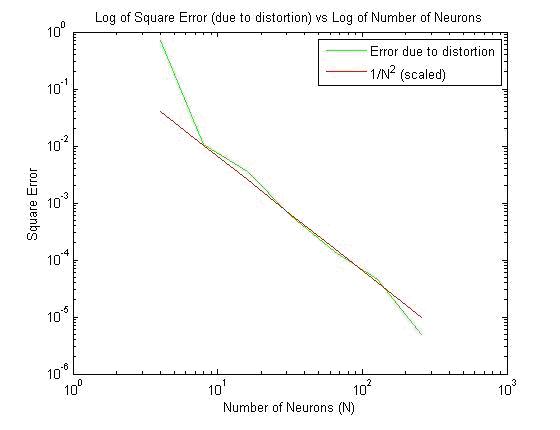}
	\caption{Error due to static distortion with increasing number of neurons.}
	\end{figure}

	The next numerical experiment performed with this model compares the above results (using Lorentzian tuning curves) to results using the Leaky Integrate-and-Fire (LIF) model and Rectified Linear Models.  

The plots of $x$ and $\hat{x}$ for these two models are omitted due to their close resemblance to the plots of Fig. 7 and Fig. 8 above.  The RMS deviation for the LIF model however, was only 0.0053; and for the Rectified Linear model, it was only 0.0027.  This suggests that the severe non-linearity in the Lorentzian tuning curves decreases the precision of their decoding of signals using optimal linear decoders.  Nevertheless, the Lorentzian tuning curves were chosen for the biological plausibility argument presented above, and since they do not perform extremely worse than LIF neurons or Rectified Linear neurons in decoding the above signal, I have chosen to continue with using tuning curves of this form for the remainder of this paper.

  The final numerical experiment performed with this model compares the above results (using Lorentzian tuning curves) where the decoders are optimal linear decoders, to results in which the decoders used were the preferred directions.  A similar approach for defining the decoders was used in Georgopoulos' model of arm movements in monkeys\cite{1984Georgopoulos}.  The result of this numerical experiment is displayed in Fig. 11, and it demonstrates that these decoders to not work very well.  The RMS deviation in this case was approximately 0.73, which is higher than that of the optimal linear decoders case by a factor of approximately 66.36.

  	\begin{figure}[h!]
	\centering		
	\includegraphics[width = \textwidth, height = 0.4 \textheight]{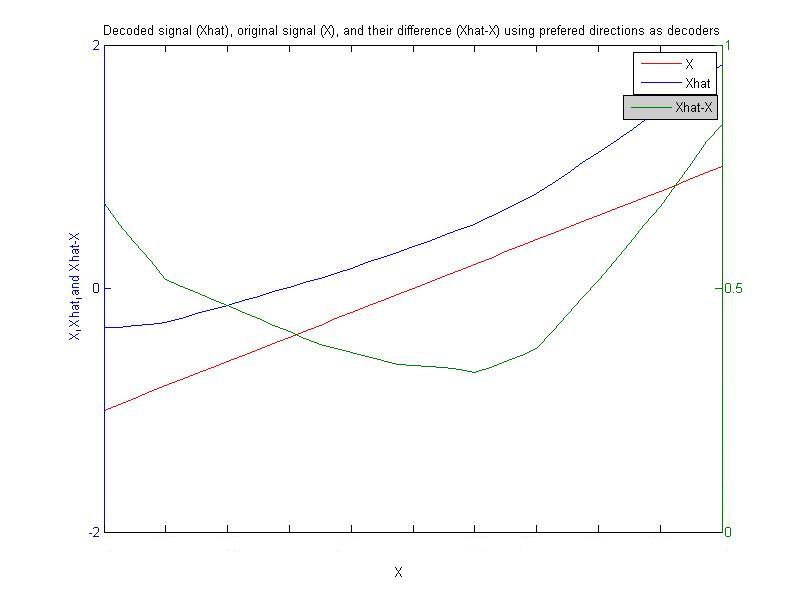}
	\caption{Decoding using preferred directions rather than optimal linear decoders.}
	\end{figure}

	In accordance with the above results, the remainder of this paper will use optimal linear decoders and Lorentzian tuning curves.

\section{1-Dimensional Representation: Equiangular Signals Varying in the Radial Dimension}

\subsection{System Description}

\subsubsection{Connectivity and Functional Functional Description}

\qquad In this section, the sound source will no longer vary in angle, but will rather vary in radial distance away from the human that is decoding it.  The sound source considered is one that emits sounds at a constant frequency and intensity (therefore, the human decoder should perceive that sound being louder when the radial distance is small, and softer when the radial distance is large).  

The brain's mechanism for detecting differences in volume (sound intensity) relies on the fact that the sound intensity entering the ipsilateral ear is greater than the sound intensity entering the contralateral ear (assuming that the head is spherical).  This is because sound intensity in the radial direction for a spherical sound source is given by\cite{2003Serway}:

\begin{equation}
I(r) = \frac{P}{4\pi r}
\end{equation}

where $I(r)$ is the intensity, $P$ is the power, and $r$ is the radial distance away from the centre of the source.  This inverse-square relationship, although defined for spherical sound sources, holds approximately true for non-spherical source's as well. As depicted in Fig. 12, for a given angle, as the sound source moves farther away from the head, the amount of space through which the sound waves travel on their way to the contralateral ear consistently remains larger than that of the ipsilateral ear, but the \textit{difference} in path lengths decreases.  This means that the intensity difference should get smaller as the sound source moves farther away from the detector's head.

  	\begin{figure}[h!]
	\centering		
	\includegraphics[height = 0.4 \textheight] {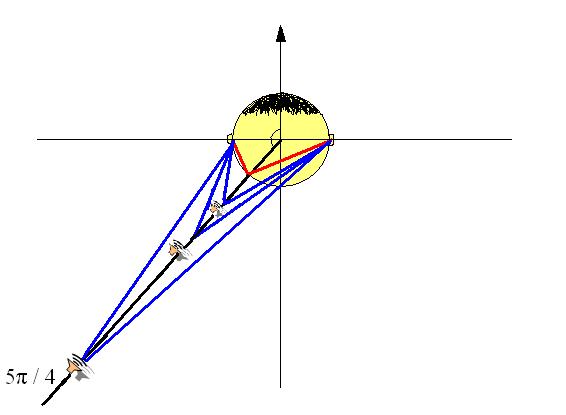}
	\caption{Sound intensity difference decreasing as the sound source is taken radially farther from the detector's head.}
	\end{figure}

On the ipsilateral side, the neurons of the posteroventral cochlear nucleus\footnote{As opposed to the angular sound localization discussed in the previous section, where the signals are instead sent from the \textit{anteroventral cochlear nucleus}.} transmit signals to the nucleus of the trapezoidal body, and neurons of the trapezoidal body send signals to the lateral superior olive.  On the contralateral side, the neurons of the cochlear nucleus transmit signals directly to the nucleus of the lateral superior olive\cite{2000Kandel}.  Notice that if the sound source was colinear with the azimuth, there would be no difference in intensity between the left and right ears (again, assuming the head is a perfect sphere).  In this case the distinction between contralateral and ipsilateral ears cannot be made, and that is why the brain first locates the sound source's angle with respect to the azimuth (as described in the above section).  Fig. 13 organizes and summarizes this information in a flow chart. 

  	\begin{figure}[h!]
	\centering		
	\includegraphics[width = \textwidth]{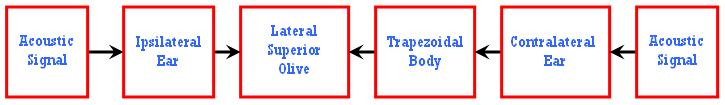}
	\caption{A flow chart depicting how the acoustic signals from the environment reach the lateral superior olive, where radial location of the sound source occurs.}
	\end{figure}

	The neurons of the lateral superior olive now have signals entering from both the ipsilateral and contralateral ears, and each neuron requires a certain amount of input current in order to be excited.  The ones that require the most input current in order to be excited are the ones that are sensitive to signals that are radially close from the detector (where the sound intensity is greatest, and therefore the neurons have evolved to ``expect'' more input current).  These neurons will be considered `off' LIF neurons, since their firing rate is expected to increase as their radial distance from the signal decreases (just like the LIF tuning curve shapes).  The ones that require the most input current will be the ones with low-valued x-intercepts, and the ones that don't require as much will be the ones with larger x-intercepts (their firing rate will still be non-zero even for low intensities and far radial distances).      

\subsubsection{Tuning Curves}

  	\begin{figure}[h!]
	\centering		
	\includegraphics[width = \textwidth, height = 0.5 \textheight]{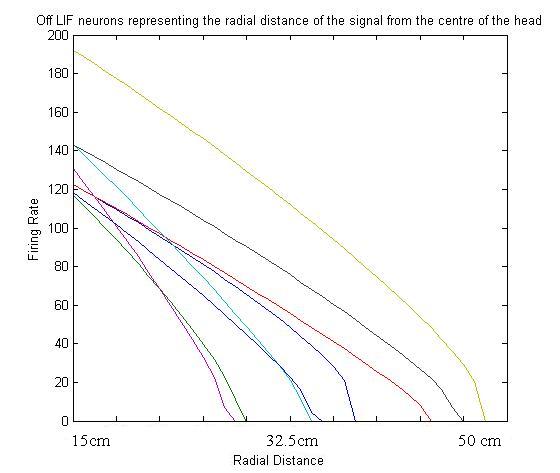}
	\caption{LIF off neurons representing radial distance from the centre of the head.}
	\end{figure}

\subsubsection{Variable Specifications and Mathematical Description of the System Function}

\qquad The variable specifications are the same as before except instead of using Lorentzian tuning curves we use `off' LIF neural response functions.  The decoding equation is the same as above.

\subsection{Design Specification}

\subsubsection{Range}

\qquad  The range for the radial distances will be 15cm to 50cm.  
\subsubsection{Signal to Noise Ratio and Precision}
\qquad As before, we assume the signal to noise ratio is 10:1, so the noise will be modeled as independent, Gaussian distributed noise with a mean of zero and a variance of 0.1.  The precision, as in the last section, will be considered the \textit{purpose} of the simulation.



\section{Feed-Forward Representation}

\qquad The first model discussed above dealt with signals about the sound source's angle with respect to the azimuth being sent through the anteroventral cochlear nuclei to the nucleus of the medial superior olive.  The second model discussed above dealt with signals about the sound source's radial distance away from the detector being sent through the posteroventral cochlear nuclei to the nucleus of the lateral superior olive.  Of course, to be able to pinpoint the location of an object, one needs information about both of these positions (both angular and radial).  This section describes the amalgamation of these two items of information.

\subsection{System Description}

\subsubsection{Connectivity and Functional Functional Description}

\qquad The final representation of the location of an object can be thought of as a 2-component vector where one component is the representation of the angular position, while the other component is the representation of the radial position.  The most important location where the representation of this information is stored is in the neurons of the primary auditory cortex\cite{2000Kandel}.  The pathway is outlined in Fig. 14.  The nuclei of the medial and lateral and superior olive send signals to the neurons of the inferior colliculus, which send signals to the medial geniculate nucleus, which sends signals to the primary auditory cortex.

 \begin{figure}[h!]
   \centering
   \subfloat{\label{fig:fullerene}\includegraphics[width = \textwidth]{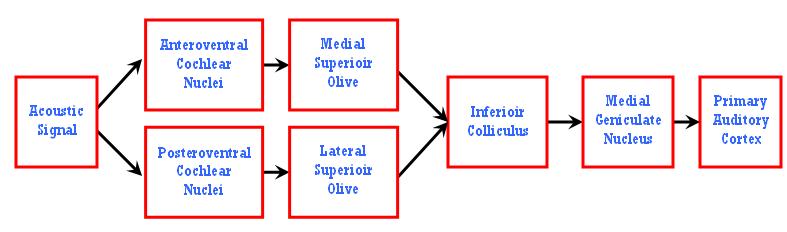}} \\
   \subfloat{\label{fig:epcot}\includegraphics[width = \textwidth]{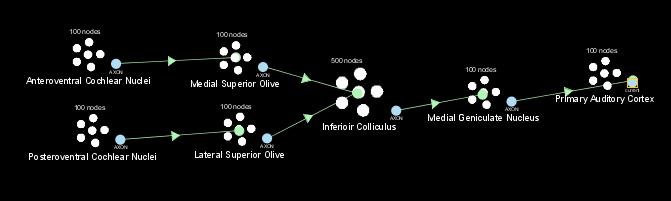}} 
   \caption{A chart displaying the flow of acoustic information through various locations in the brain on its way to the primary auditory cortex.  The lower figure is a screenshot from the neural engineering software NENGO that was used for this element of the simulation (NENGO stands for Neural Engineering Objects: www.nengo.ca)}
   \label{fig:fullerenes}
 \end{figure}



In the network displayed above, the representation of a signal by one population of neurons, serves as the signal to the next population.  Then the representation of that representation, serves as the signal for the next population, and so on.

\subsubsection{Tuning Surfaces}

\qquad As described in \cite{2000Kandel}, specific neurons represent specific spatial locations of sound sources (both angular and radial), so certain neurons will have high firing rates when the sound source is a certain angle and distance away, and should have low firing rates otherwise.  For this reason, the tuning surfaces were taken to be 2-dimensional Gaussian functions, with peak firing rates chosen to be between 114 and 268 spikes/second (as described in section 1).  
	An example of one of these tuning surfaces is shown in Fig. 16.  It's peak is found at approximately (x,y) = (0.25,0.56) and is therefore most sensitive to sound sources at that location:
	
\begin{figure}[h!]
	\centering		
	\includegraphics[width = \textwidth , height = 0.75 \textheight]{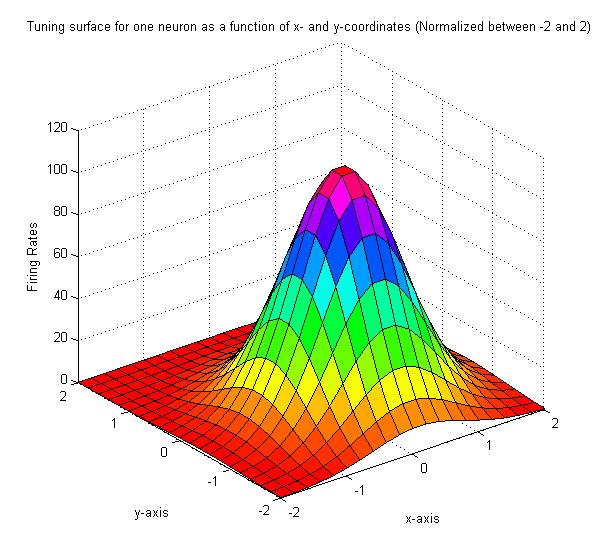}
	\caption{The tuning surface for one neuron whose preferred location is approximately (x,y) = (0.25,0.56) .}
	\end{figure}
	
Fig. 17 displays tuning surfaces for 15 different neurons, all with different preferred locations.

\begin{figure}[h!]
	\centering		
	\includegraphics[width = \textwidth , height = 0.75 \textheight]{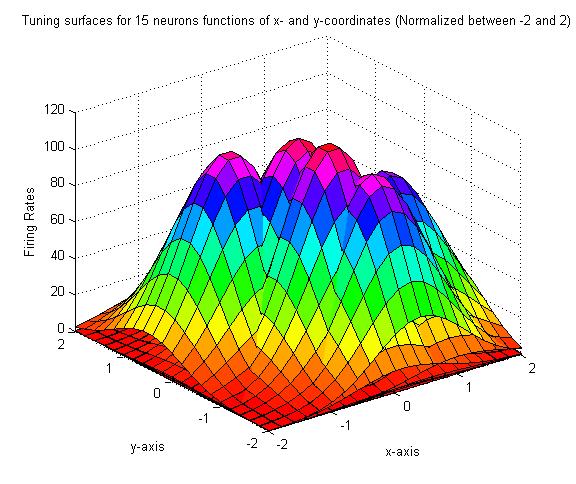}
	\caption{The tuning surface for 15 neurons whose preferred locations are randomly distributed between over the plane: $-2\le x\le 2 , -2\le y \le 2$.}
	\end{figure}
	
\subsubsection{Variable Specifications and Mathematical Description of the System Function}
	
	\qquad For this model, the variables x and y will represent spatial locations in a 2-Dimensional flat plane (parallel to the detector's transverse plane).  

	The encoding of these variables by each neuron will be expressed as 2-dimensional Gaussian tuning surfaces given by the neural response functions (as depicted in Figs. 16 and 17):
	
	\begin{equation}
	a_i(x,y) = FRe^{-(x-x_i)^2-(y-y_i)^2}
	\end{equation}
	
	where FR is the peak firing rate (picked between 114 and 268 spikes/second) and $x_0$ and $y_0$ are the coordinates of the preferred locations of the $i$th neuron (each chosen between -2 and 2).

\subsection{Design Specification}

\subsubsection{Range}

\qquad The range of x (the angular component running from 0 to $2\pi$ radians) and y (the radial component running from 15cm-50cm) are both normalized to be between -2 and 2.

\subsubsection{Signal-to-Noise Ratios (SNR) and Precision}

\qquad As before, we assume that the signal-to-noise ratio is approximately 10:1, and therefore the noise will be independent, Gaussian distributed, and with a mean of 0 and variance of 0.1 for each neuron in the model.  Likewise, the precision of the decoding will be one of the \textit{purposes} of the simulations.











\section{Extrapolation: Spatially-Varying Sound Velocities}

\qquad  The neurons of our primary auditory cortex have evolved to represent specific locations for sound sources.  So if the sound source is at one particular location, there are particular neurons that are most sensitive to the stimulus that will fire, whereas if the sound source was moved, different neurons would be more sensitive.  These neurons are getting their signals from the neurons of the superior olivary complex, who (in the case of detecting angular positions of sound sources) fire based on certain time differences between the signal reaching different ears.  A hypothetical example would be, if the right ear attained the signal 500 $\mu s$ before the left ear, it might mean that the angle the sound source made with the azimuth was $\frac{\pi}{4}$ radians, and the neurons of the superior olivary complex would send signals to the primary auditory cortex informing them that the signal was coming from that direction.  

This will cause problems if the left ear is in water and the right ear is in air, as in the case of someone floating sideways in a swimming pool.  This is because the speed of sound in water is slower than the speed of sound in air, so the time difference might be delayed even more than usual.  However, the neurons of the primary auditory cortex (as far as we understand them) do not ``know'' that one ear is in water and one is in air, so they will take this extended time difference to dictate that the sound source was at a greater angle away from the detector, while in reality the angle could have been smaller.  

Now that I have constructed a model that is able to locate the position of a sound source to a reasonable degree of accuracy, where the sound source is assumed to be of constant frequency and intensity, and is assumed to emit sound waves that are transmitted through a uniform medium (such as air), this model is able to predict the errors that the brain will make when attempting to locate signals when one ear is in water and one ear is in air.  

The speed of sound in a fluid is given by:

\begin{equation}
v = \sqrt{\frac{K}{\rho}}
\end{equation}

where $\rho$ is the density of the fluid and $K$ is the bulk modulus of the fluid\cite{2003Serway}.

Here we assume that the speed of flow of the water is much less than the speed of sound in the water (so that the additive resultant velocity of sound is not much different from the speed of sound itself). Suppose that the density of the water is 1$\frac{g}{cm^3}$
and that the Bulk Modulus is $1.056$x$10^9 \frac{kg}{s^2 cm}$ so that the speed of sound in the fluid is approximately 325$\frac{m}{s}$ as opposed to 343$\frac{m}{s}$ in air.  This means, assuming a spherical head of diameter 10cm, and a sound source making a $\frac{3\pi}{4}$ radian angle with the azimuth and 15cm away from the centre of the head, the  length of the path of the sound to the left ear will be approximately 20.98cm and to the right ear would be 11.05cm (by the Law of Cosines).  We can calculate using similar triangles that on its way to the left ear, the sound travels approximately 11.85cm of the 20.98cm path in water, and the remaining 9.13cm in air.  Using 32500cm/second for water and 34300cm/second for air, we get a total time of 0.63$\mu s$, whereas for the right ear the path is completely in water, so the time it takes is 0.34$\mu s$.  This temporal difference of 0.29$\mu s$ is in normal circumstances caused by the source being at an angle of $\frac{9 \pi}{11}$ to the azimuth (assuming its radial distance is still 15cm).  This is depicted in Fig. 18 below.  The blue dot at $\frac{3 \pi}{4}$ depicts the actual sound source and the red dot is the human's approximation of it.  

Similarly, several other errors are calculated and included in the same figure.

 	\begin{figure}[h!]
	\centering		
	\includegraphics{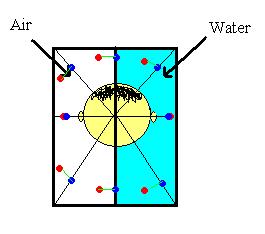}
	\caption{Errors made by the brain in detecting locations of sounds.  The sound sources are denoted in blue and are 15cm radially outward from the centre of the spherically shaped head.  The head's diameter is 10cm.  The red dots are the brain's (incorrect) estimate of the location from which the sound is arriving.}
	\end{figure}

\end{document}